# Topological Lifshitz transition-induced bipolarity of anomalous Nernst effect in kagome magnet $YCo_3$

Sheng Xu,[1,2,*] Yue-Yang Wu,[2] Hao-Ran Bai,[2] Zheng Li,[2] Shu-Xiang Li,[2] Jun-Jian Mi,[2] Tian-Hao Li,[2] Ze-Wei Wang,[2] Ze-Kai Dong,[2] Jiang Ma,[2] Xiao-Bo Wu,[1] Qian Tao,[2] and Zhu-An Xu[2,3,4,†]

[1]School of Physics, Zhejiang University of Technology, Hangzhou 310023, China

[2]State Key Laboratory of Silicon and Advanced Semiconductor Materials & School of Physics, Zhejiang University, Hangzhou 310058, China

[3]Hefei National Laboratory, Hefei 230088, China

[4]Collaborative Innovation Center of Advanced Microstructures, Nanjing University, Nanjing, 210093, China

The kagome lattice, renowned for hosting topological band structures and rich magnetic behaviors, offers an exceptional setting to investigate unconventional transport in magnetic topological systems. Controlling the polarity of the anomalous Nernst effect (ANE) is crucial for designing flexible thermoelectric devices, such as thermopiles, where the ability to switch the thermoelectric voltage sign can dramatically enhance energy conversion efficiency and output. Here, we demonstrate such a bipolar ANE in the kagome magnet $YCo_3$, driven by a temperature-induced topological Lifshitz transition. With a Curie temperature $T_C$≈225 K, sizable anomalous Hall and Nernst effects emerge below $T_C$. Supported by the first-principles calculations, the AHE and ANE are suggested to be dominated by the intrinsic mechanism. Furthermore, the intrinsic anomalous Hall conductivity exhibits a piecewise-linear dependence on magnetization, with an abrupt slope change near 100 K, consistent with the Karplus–Luttinger mechanism. Concurrently, the anomalous Nernst coefficient $S^A yx$ reverses its sign around the same temperature, realizing the crucial bipolarity. These anomalies could be interpreted as a topological Lifshitz transition, enabled by the evolution of Co moments that could shift the Fermi level relative to Weyl nodes. Our work reveals $YCo_3$ as a prototypical kagome magnet where temperature and magnetism directly govern both Weyl node topology and the bipolar ANE, opening a pathway to magnetically control thermoelectric output in topological quantum materials.

**Correspondence:** Sheng Xu (shengxu@zjut.edu.cn) Zhu-An Xu (zhuan@zju.edu.cn)

## Introduction

The thermoelectric voltage generation in solids occurs in two distinct geometries: longitudinally as the Seebeck effect, or transversely as the anomalous Nernst effect (ANE) [1–4]. The transverse configuration of ANE offers a fundamental advantage for device integration, enabling more efficient heat-sink designs since the electrical contacts are perpendicular to the heat flow. Moreover, the ability to control the polarity of the transverse voltage–akin to combining p-type and n-type legs in a conventional thermopile–is essential for constructing circuits that cumulatively enhance power output [1]. Achieving such bipolarity within a single material thus represents a key challenge in thermoelectric research [5]. Figure 1a,b displays the schematic diagrams of ANE thermopiles.

Magnetic topological materials, which host non-trivial band structures and broken time-reversal symmetry, provide an ideal platform for manipulating the ANE and the anomalous Hall effect (AHE) [6–19]. Their anomalous transverse signals stem from the Berry curvature, which can be tuned by shifting the Fermi level $E_F$ across topological band crossings via external parameters such as chemical doping or pressure [5, 20–22]. Among various topological systems, kagome magnets stand out as a particularly fertile family, unifying frustrated lattice geometry, topological electronic states, and diverse magnetic orders. Notable examples include Mn-based kagome magnets like $Mn_3Sn$ and $RMn_6Sn_6$ (R = rare-earth element) [13, 23–30], as well as Fe-based systems such as $Fe_3Sn_2$ [31–33]. A prominent Co-based kagome Weyl magnet, $Co_3Sn_2S_2$, exhibits a strong coupling between the AHE and the Co magnetic moment, as the Weyl node separation is directly governed by magnetization [10–12]. Significantly, chemical doping in $Co_3Sn_2S_2$ has been shown to tune $E_F$ across Weyl nodes, leading to sign reversal of Nernst voltage, which demonstrates the realization of bipolar ANE via Fermi level engineering [5]. Beyond chemical doping, temperature could serve as a clean and continuous tuning parameter to drive electronic transitions in topological materials. For instance, temperature-induced Lifshitz transitions have been reported in the systems such as $W_2As_3$ and $YbMnSb_2$ [34, 35], modifying the Fermi surface topology without introducing extrinsic disorder.

Inspired by this approach, we investigate the kagome magnet $YCo_3$ as a platform to probe the interplay among temperature, magnetic order, and anomalous transverse transport. The Curie temperature $T_C$ of our sample is about 225 K, lower than previously reported values (~301–328 K) [36–40]. We attribute this reduction to minimal La substitution at Y sites during crystal growth, which modifies Co–Co distances and weakens the direct exchange interaction. Below $T_C$, we observe pronounced intrinsic AHE and ANE, which are supported by the firstprinciples calculations. Notably, the intrinsic anomalous Hall conductivity (AHC) $\sigma^A_{yx}$ shows a piecewise-linear dependence on magnetization, consistent with the Karplus–Luttinger (K–L) mechanism [41], but exhibits a distinct slope change near 100 K. Concurrently, the anomalous Nernst coefficient $S^A_{yx}$ undergoes a sign reversal around the same temperature. These correlated anomalies suggest a temperature-driven Lifshitz transition, likely mediated by the enhancement of Co magnetic moments that enlarges band splitting and thereby modifies the Weyl node configuration, leading to the sign reversal of the Berry curvature.

Our study establishes $YCo_3$ as a model kagome magnet in which a temperature-induced topological Lifshitz transition directly controls the Berry curvature and ANE polarity. In contrast to the chemical doping demonstrated in Co3Sn2S2, this work highlights temperature as a continuous and clean tuning parameter for achieving bipolar ANE, offering a distinct approach for developing topological thermoelectric devices.

## Experimental Details

Plate-like $YCo_3$ single crystals were grown by the flux method. A stoichiometric mixture of Y:Co:La (1:3:2 molar ratio) was sealed in an alumina crucible within a quartz ampoule under vacuum. The ampoule was heated to 1430 K, homogenized for 10 h to ensure complete melting and mixing, then slowly cooled at a rate of 3 K/h to 1170 K. Excess molten La flux was decanted via centrifugation at 1170 K. The atomic composition of the obtained single crystal was measured by energy dispersive x-ray spectroscopy (EDX). Single-crystal X-Ray diffraction (XRD) was carried on a Bruker D8 Venture diffractometer with Mo Kα radiation. Apex5 was employed for the XRD refinement.

The measurements of transport properties were performed on Quantum Design physical property measurement system (QD-PPMS). The measurements of magnetic properties were performed on Quantum Design magnetic property measurement system (QD-MPMS). The longitudinal and transverse resistivity was collected by standard six-probe method. Thermal gradient was established by a 1 kΩ resistor heater in the measurements of thermoelectric properties. The Nernst and Hall signals were measured at both positive and negative magnetic field polarities, and the difference of the two polarities was deduced to remove the longitudinal contribution from the voltage probe misalignment.

The first-principles calculations were performed by using Vienna Ab-initio Simulation Package (VASP) [42, 43] based on the density functional theory (DFT) [44, 45]. The generalized gradient approximation (GGA) of Perdew–Burke–Ernzerhof (PBE) type [46] was chosen for the exchange-correlation functional. The kinetic energy cutoff of the plane-wave basis was set to be 500 eV. A $7 \times 7 \times 7$ *k*-point mesh was used for the Brillouin zone (BZ) sampling.

## Results and Discussion

Figure 2(a) presents the crystal structure of the intermetallic compound $YCo_3$, which crystallizes into a layered structure with symmetry $R\bar{3}m$ (No. 166). Its lattice parameters are reported to be a = b = 4.992 Å, and c = 24.300 Å [40], respectively. The unit cell contains three formula units and five crystallographically inequivalent atomic sites, labeled as Y-1, Y-2, Co-1, Co-2, and Co-3. Notably, the Co-1 atoms form a two-dimensional kagome lattice parallel to the ab-plane (Fig. 2(b)). Figures 2(c) and 2(d) show the single-crystal XRD patterns of $YCo_3$ for the (0kl) and (hk0) reciprocal lattice planes, respectively. Rietveld refinement confirms that the crystal structure adopts the R3m space group. The refined lattice parameters, a = b = 5.049(2) Å, and c = 24.393(5) Å, are slightly larger than previously reported values. This subtle lattice expansion is likely due to the partial substitution of La at the Y sites during crystal growth, which is consistent with our EDX results indicating a La doping level of approximately 5-7 %.

Figure 3(a) shows the temperature-dependent resistivity of $YCo_3$. The resistivity decreases with decreasing temperature, consistent with metallic behavior. A broad kink near 225 K suggests the onset of a magnetic phase transition. The upper inset shows an optical image of the as-grown single crystal, and the lower panel displays the first derivative of resistivity with respect to temperature. The derivative indicates a broadened phase transition occurring between approximately 207 K and 225 K. To further investigate the magnetic properties, we performed detailed magnetic susceptibility measurements. As shown in Fig. 3(b), the compound exhibits a strong magnetic anisotropy. For a magnetic field B = 0.1 T applied along the c-axis, the magnetization shows a rapid increase below 225 K upon cooling, followed by a plateau, characteristic of a ferromagnetic transition. For the field within the ab-plane, the transition occurs near 207 K, with a significantly weaker magnetic response, indicating that the c-axis is the easy magnetization direction. The transition temperatures, determined from minima in the first derivative of magnetization with respect to temperature (Fig. 3(c)), agree well with the anomalies observed in resistivity. A similar anisotropic behavior was reported in the parent compound $YCo_3$, which exhibits Curie temperatures of 328 K ($B$ // $c$) and 294 K ($B$ // $ab$) [47]. Note that the $T_C$ of $YCo_3$ is highly sensitive to slight changes in Co stoichiometry, and prior studies have shown that doping on the Co sites can strongly affect the magnetic transition [48, 49]. Moreover, the magnetic properties of $RCo_3$ (R=rare-earth element) compounds also exhibit considerable variation depending on the specific rare-earth atom occupying the R-site [38, 40, 50–52].. Thus, the reduced $T_C$ in our sample is attributed to lattice expansion due to La substitution, which should increase the Co-Co bond distances and weaken the ferromagnetic direct exchange interaction. This conclusion is supported by the lattice parameter changes obtained from single-crystal XRD refinement. Figure 3(d) displays the magnetic field dependence of the magnetization at various temperatures for B // ab. Below 220 K, the magnetization increases with increasing applied field and eventually tends to saturate.

Figure 4(a) presents the magnetization curve with the magnetic field applied along the *c*-axis, which exhibits typical ferromagnetic behavior with an increase in saturation magnetization as temperature decreases. At 10 K, the saturation magnetization is measured to be 1.32$\mu_B$/f.u., which corresponds to 0.44$\mu_B$/Co. This value is slightly lower than those in previous reports, which range between 0.5$\mu_B$/Co and 0.7$\mu_B$/Co [39, 53]. The reduction in the moment is consistent with the lattice expansion and modified exchange interactions induced by La doping. Figure 4(b) displays the field-dependent Hall resistivity measured from 10 K to 300 K with the current applied in ab-plane and the field along *c*-axis. Below $T_C$, the Hall resistivity increases rapidly in the low-field region and subsequently tends to saturate, demonstrating a clear signature of the AHE. As the temperature decreases, the value of the Hall resistivity plateau increases. At 10 K, the Hall resistivity plateau reaches a maximum value of approximately 1.92 μΩ · cm. The Hall resistivity is comprised of both ordinary ($\rho^O_{yx}$) and anomalous ($\rho^A_{yx}$) contributions. Thus, the total Hall resistivity can be expressed as:

$$\rho_{yx} = \rho^O_{yx} + \rho^A_{yx} = R_0B + R_S\mu_0M \quad (1)$$

where $R_0$ is the ordinary Hall coefficient, $R_S$ is the anomalous Hall coefficient, and $M$ denotes the magnetization. By performing a multivariate linear fitting to the Hall resistance using Equation (1), we obtain the value of $R_0$, as shown in Figure 4c. It can be seen that $R_0$ gradually increases with increasing temperature, and tends to be nearly temperature-independent above 100 K.

Empirically, the mechanism of the AHE can be classified into three distinct regimes based on the relationship between the Hall conductivity and the longitudinal conductivity: (1) the dirty regime, where the longitudinal conductivity is less than $3\times10^3$ $\Omega^{-1}$cm$^{-1}$;(2)the intrinsic Berry-curvature-dominated regime; and (3) the clean regime, characterized by a longitudinal conductivity exceeding $5\times10^5$ $\Omega^{-1}$cm$^{-1}$. For the $YCo_3$ single crystal, the measured longitudinal conductivity ranges from $8.8\times10^3$ to $1.1\times10^5$ $\Omega^{-1}$cm$^{-1}$ between 2.5 and 300 K, placing it within the intrinsic regime. Combined with the presence of a Co-based kagome lattice, these results suggest a possible contribution to the AHE from the intrinsic nonzero Berry curvature, arising from the topological electronic band structure. The dominant mechanism of the AHE can be discerned by analyzing the relationship between the anomalous Hall resistivity $\rho^A_{yx}$ and the longitudinal resistivity $\rho_{xx}$ using the expression:

$$\rho^A_{yx} = a(M)\rho_{xx} + b(M)\rho^2_{xx} \quad (2)$$

The linear term in $\rho_{xx}$ corresponds to skew scattering, with $a(M)$ proportional to the $M$, while the quadratic term corresponds to intrinsic or side-jump mechanisms, where $b(M)$ relates directly to the intrinsic AHC which also exhibits linear scaling with $M$ [54, 55]. Thus, the intrinsic and extrinsic contributions can be separated by linearly fitting $\rho^A_{yx}/(M\rho_{xx})$ against $\rho_{xx}$. Interestingly, as shown in Fig. 4(d), the plot of $\rho^A_{yx}/(M\rho_{xx})$ versus $\rho_{xx}$ exhibits two distinct linear dependencies with different slopes, separated at approximately 100 K. This linear behavior is consistent with the K-L mechanism [41]. After subtracting the skew-scattering contribution, the intrinsic AHC is displayed in Fig. 4(e). At 10 K the AHC is about 580 $\Omega^{-1}\cdot$ cm$^{-1}$ and decreases with increasing temperature, before undergoing an abrupt change near 100 K. To evaluate the side-jump mechanism's role in the AHE, we employ the established relation $\sigma_{xy} = -\frac{e^2}{\hbar}\int_{BZ}\frac{d^3k}{(2\pi)^3}f(k)\Omega_B(k)$

$\frac{e^2}{ha}(\frac{\varepsilon_{SO}}{E_F})$, in which $\varepsilon_{SO}$ corresponds to the spin-orbit coupling energy [56]. Applying this relation with a lattice parameter of a~ $V^{1/3}$ = 8.46 Å and a characteristic energy ratio $\varepsilon_{SO}/E_F$~0.01 for metallic ferromagnets, the computed side-jump conductivity is found to be roughly 4.6 $\Omega^{-1}\cdot$ cm$^{-1}$. This result indicates that the extrinsic side-jump process contributes only minimally to the overall AHC, particularly in comparison to the intrinsic component. Moreover, as shown in Fig. 4(f), $\sigma^{A,\ in}_{xy}$ also varies linearly with magnetization, with a noticeable anomaly near 100 K. This scaling behavior of the intrinsic AHC with magnetization closely resembles that observed in $Co_3Sn_2S_2$, where Weyl nodes shift monotonically with the constrained magnetic moment on Co atoms, modifying the Berry curvature [10]. Since the intrinsic AHC is given by the jump in $\sigma^{A,\ in}_{xy}$ near 100 K is likely attributed to a redistribution of the Berry curvature driven by a temperature-induced topological Lifshitz transition.

Given that the intrinsic AHC captures the integral Berry curvature over all occupied bands and the anomalous Nernst signal is mainly sensitive to the Berry curvature near the Fermi level [13, 57], we carried out systematic measurements of transverse thermoelectric response to further investigate the effect of the possible temperature-induced topological Lifshitz transition on ANE. Figure 5(a) displays the fielddependent Nernst effect at different temperatures with obvious ANE below $T_C$ . Interestingly, the anomalous Nernst thermopower exhibits a negative signal below 100 K and switches to positive above 100 K. About 40 K, the anomalous Nernst thermopower reaches a maximum value of approximately 0.98 μV K$^{-1}$. This value is smaller than that of Co-based kagome magnets $Co_3Sn2S_2$ [12] and $LaCo_5$ [58], and comparable to the Mn-based kagome magnets $Mn_3Sn$ [13] and $TbMn_6Sn_6$ [30]. To gain deeper insight into the intrinsic ANE, we extract the Nernst conductivity using the relation:

$$\alpha_{yx} = \sigma_{xx}S_{yx} + \sigma_{yx}S_{xx} \tag{3}$$

where $\sigma_{xx}$ is the longitudinal conductivity and $S_{xx}$ is the thermopower, respectively. Figure 5(b) presents the measured isothermal field-dependent Nernst conductivity. The anomalous Nernst conductivity (ANC), defined as $\alpha^A_{yx} = \sigma_{xx}S^A_{yx} + \alpha^A_{yx}S_{xx}$, is subsequently derived and also displayed in Fig. 5(d). Similarly, the anomalous Nernst conductivity also undergoes a sign reversal near 100 K, with negative values below and positive values above this temperature. Figure 6 presents a comparative scatter plot of the anomalous Nernst thermopower $|S^A_{yx}|$ as a function of magnetization $M$ across various magnetic topological materials. The value for $YCo_3$ lies outside the typical range of conventional ferromagnets, providing evidence for the dominant role of an intrinsic Berry curvature mechanism.

As illustrated in Fig. 5(c) and analogous to the polarity control of the Seebeck coefficient in semiconductors [1], the relative position between the Fermi level and the Weyl nodes is the key parameter that governs this Berry curvature distribution, thereby directly controlling the polarity of the ANE [5]. Figure 5(d) displays the temperature dependence of both the anomalous Nernst thermopower, the ANC, and the Seebeck coefficient. Notably, the sign reversal of the anomalous Nernst signal occurs near the temperature where the Seebeck coefficient approaches its maximum value. Given the direct relationship between the Seebeck coefficient and the nature of charge carriers, this correlation provides further evidence supporting the possible occurrence of a Lifshitz transition around 100 K. As illustrated in Fig. 3(b), the spontaneous magnetization of Co could increase with decreasing temperature, likely leading to an increased energy splitting between two originally degenerate bands. This change may shift the relative positions between the Weyl nodes and the Fermi level, thereby modifying the distribution of Berry curvature, which is also discussed in $Co_3Sn_2S_2$ [10, 11]. Given that the anomalous Nernst effect is highly sensitive to Berry curvature near the Fermi energy, it is therefore proposed that such a shift accounts for the observed reversal in the thermopower polarity.

According to the Topological Magnetic Materials database [61-64], $YCo_3$ is considered likely to host Weyl states near the Fermi level. To further investigate its electronic structure, we performed first-principles calculations. The band structure with spin-orbit coupling (SOC) in the ferromagnetic ground state, obtained from our DFT calculations, is shown in Fig.7(a). The results indicate that the electronic states near the Fermi level originate predominantly from Co atomic orbitals–especially from the Co-1 atoms that form the kagome lattice–with only minor contributions from the Y-1 and Y-2 sites, as detailed in the partial density-of-states (pDOS) analysis in Fig.7(b). Notably, the features suggestive of band crossings appear in close proximity to the Fermi level. For example, the crossings A and $B^-$ along the Γ-T direction are located approximately 5 meV and 4 meV from the Fermi level, respectively. By calculating the position of the band node and the chirality associated with each node, we determined that the point $B^-$ carries a chiral charge of −1, which unambiguously identifies it as a Weyl point. Its time-reversal counterpart, the point $B^+$, carries a chiral charge of +1, forming a Weyl pair with the point $B^-$. Further experimental verification, such as Angle-Resolved Photoemission Spectroscopy measurements would be valuable to confirm the existence and nature of these crossings. Finally, while our study on lightly La-doped $YCo_3$ reveals a clear signature of a temperature-driven topological Lifshitz transition, we note that such a Lifshitz transition temperature in the pure compound could be changed, as inferred from the sensitivity of the near-Fermi-level band crossings to external perturbations.

## CONCLUSION

In conclusion, we have systematically investigated the anomalous transverse transport in the kagome magnet $YCo_3$, which exhibits a ferromagnetic transition at $T_C \approx 225$ K. Pronounced intrinsic anomalous Hall and Nernst effects are observed below $T_C$. A key finding is the piecewise-linear dependence of the intrinsic anomalous Hall conductivity on magnetization, accompanied by a distinct slope change near 100 K. Concurrently, the anomalous Nernst coefficient undergoes a sign reversal around the same temperature, being negative below and positive above, thereby realizing a bipolar ANE within a single material. The concurrence of these features—namely, the slope change in $\sigma^{A,in}{}_{xy}(M)$, the sign reversal in $S^{A}{}_{yx}$, and the correlated anomalies in the Seebeck coefficient—along with first-principles calculations revealing multiple band crossings extremely close to the Fermi energy, provides compelling evidence for a temperature-driven topological Lifshitz transition. This transition is likely attributed to the evolution of Co magnetic moments with temperature, which may modify the band splitting and consequently shift the Fermi level relative to the Weyl nodes, potentially leading to a reconstruction of the Berry curvature distribution.

Our work establishes $YCo_3$ as a prototypical kagome magnet in which temperature and magnetism directly govern the Weyl node topology and Berry curvature, thereby enabling a thermally tunable bipolar anomalous Nernst effect (ANE). This tunability provides the foundational element for developing ANE-based thermopiles, paving the way for highperformance, flexible energy conversion devices.

Acknowledgements
We thank Dr. Jian-Feng Zhang for the valuable discussions on the band structure. This work was supported by the National Natural Science Foundation of China (Grant No. 12574177; 12174334; 12204410), the Innovation program for Quantum Science and Technology (Grant No.2021ZD0302500), the Natural Science Foundation of Zhejiang Province (LMS25A040002).

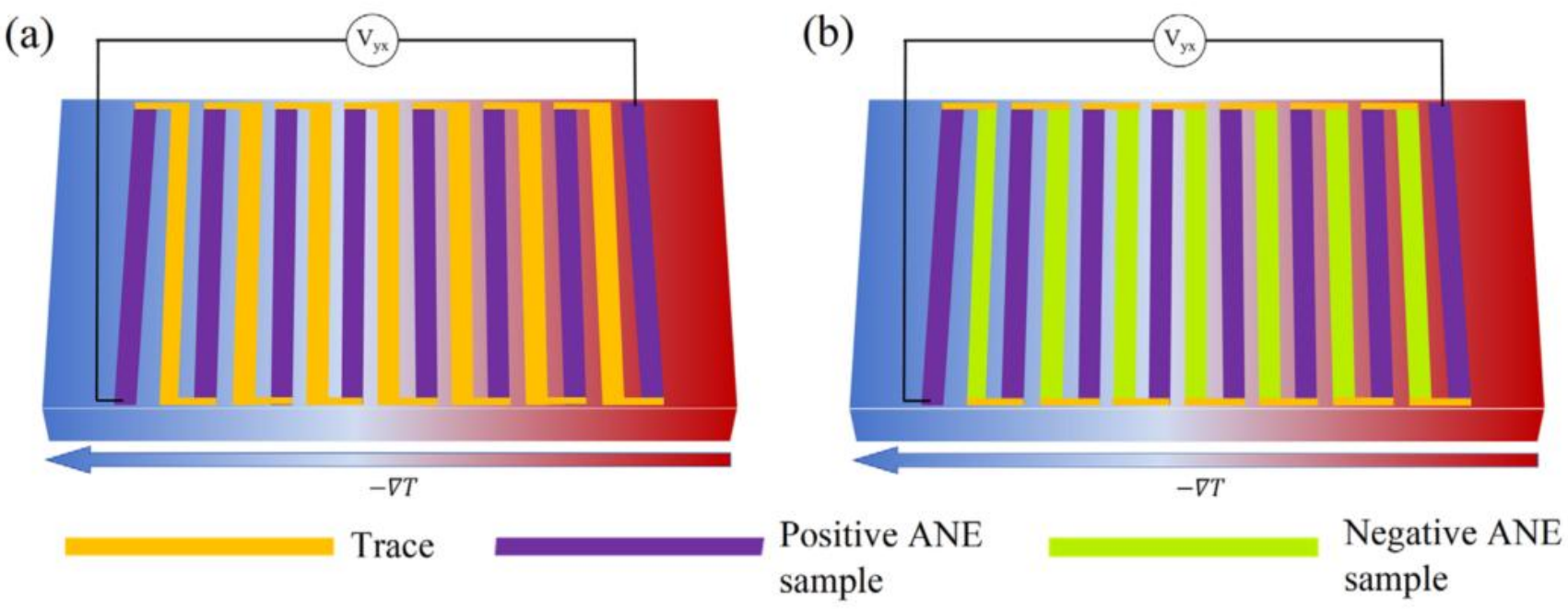

FIGURE 1 (a) and (b) Schematic diagrams of anomalous Nernst effect thermopiles. The bipolar design in (b) increases the thermopile density and transverse voltage output.

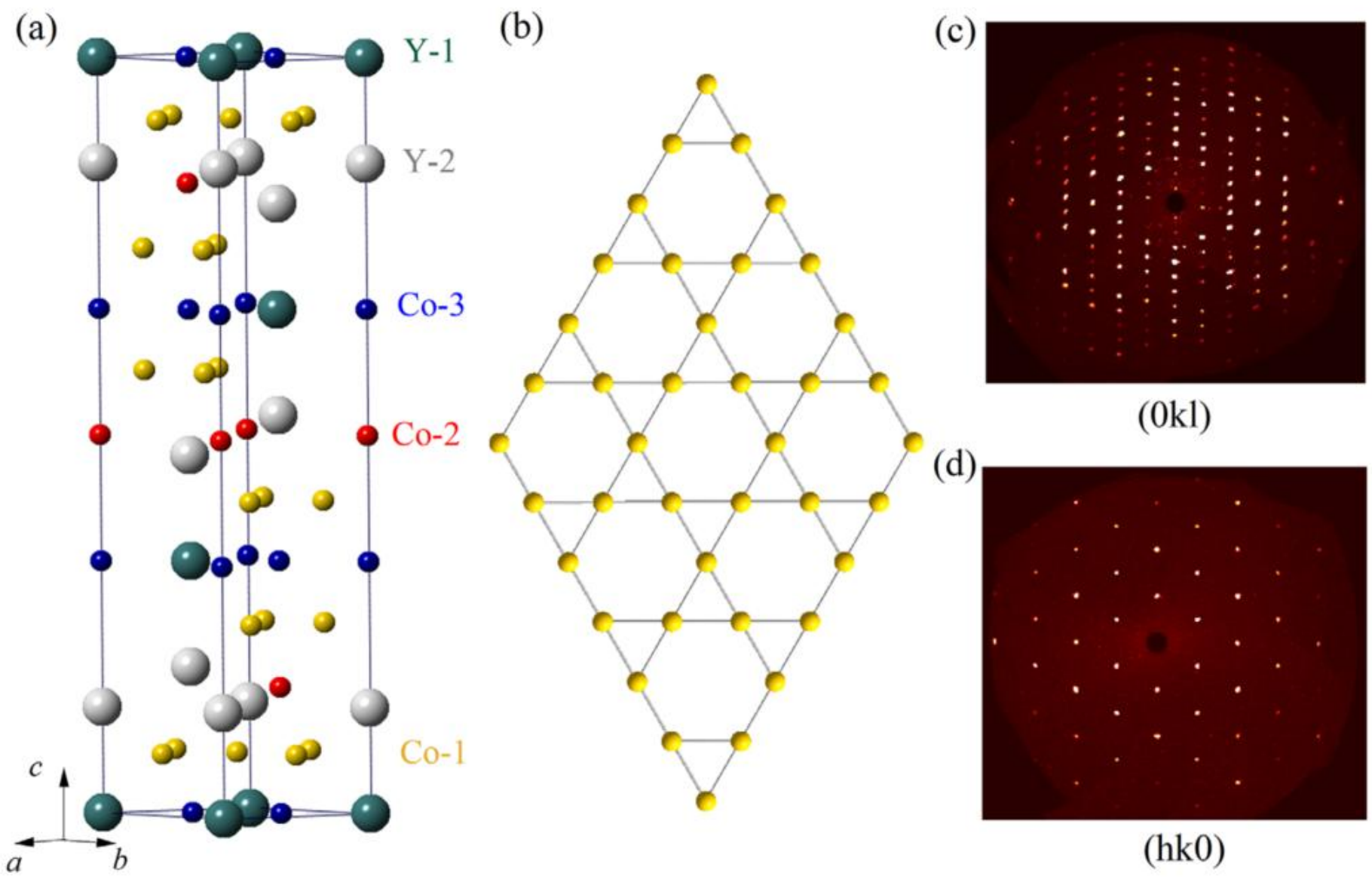

FIGURE 2 (a) Crystal structure of $YCo_3$. (b) Co-1 kagome layer in $YCo_3$. (c) and (d) Single XRD patterns of (0kl) and (*h*hk0) planes, respectively.

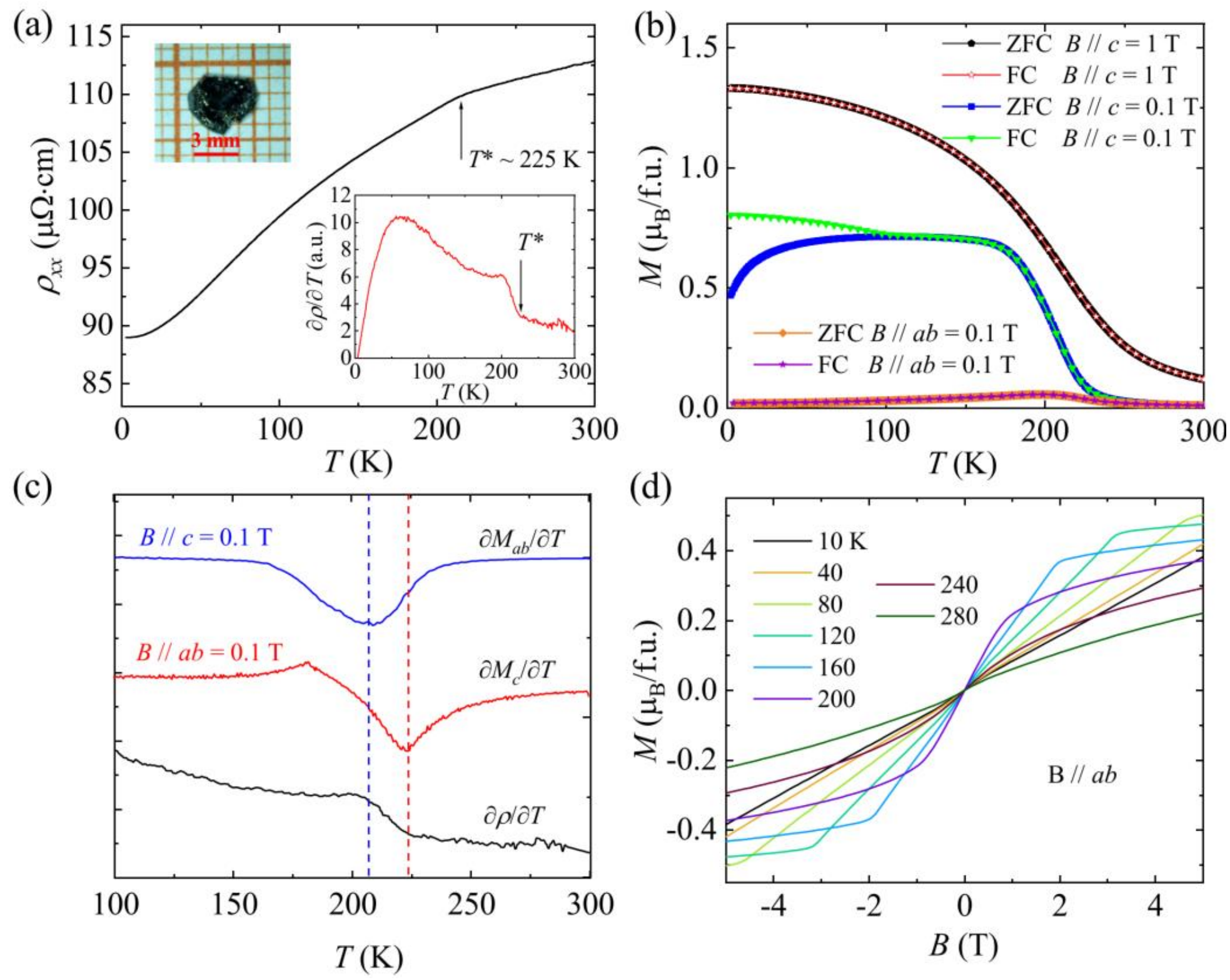

FIGURE 3 (a) Temperature-dependent resistivity from 2.5 to 300 K. Insets show the picture of the as-grown crystal and the temperaturedependent $\partial\rho_{xx}/\partial T$. (b) Temperature-dependent magnetization. (c) First derivative of resistivity and magnetization with respect to temperature. (d) Field-dependent magnetization at various temperatures.

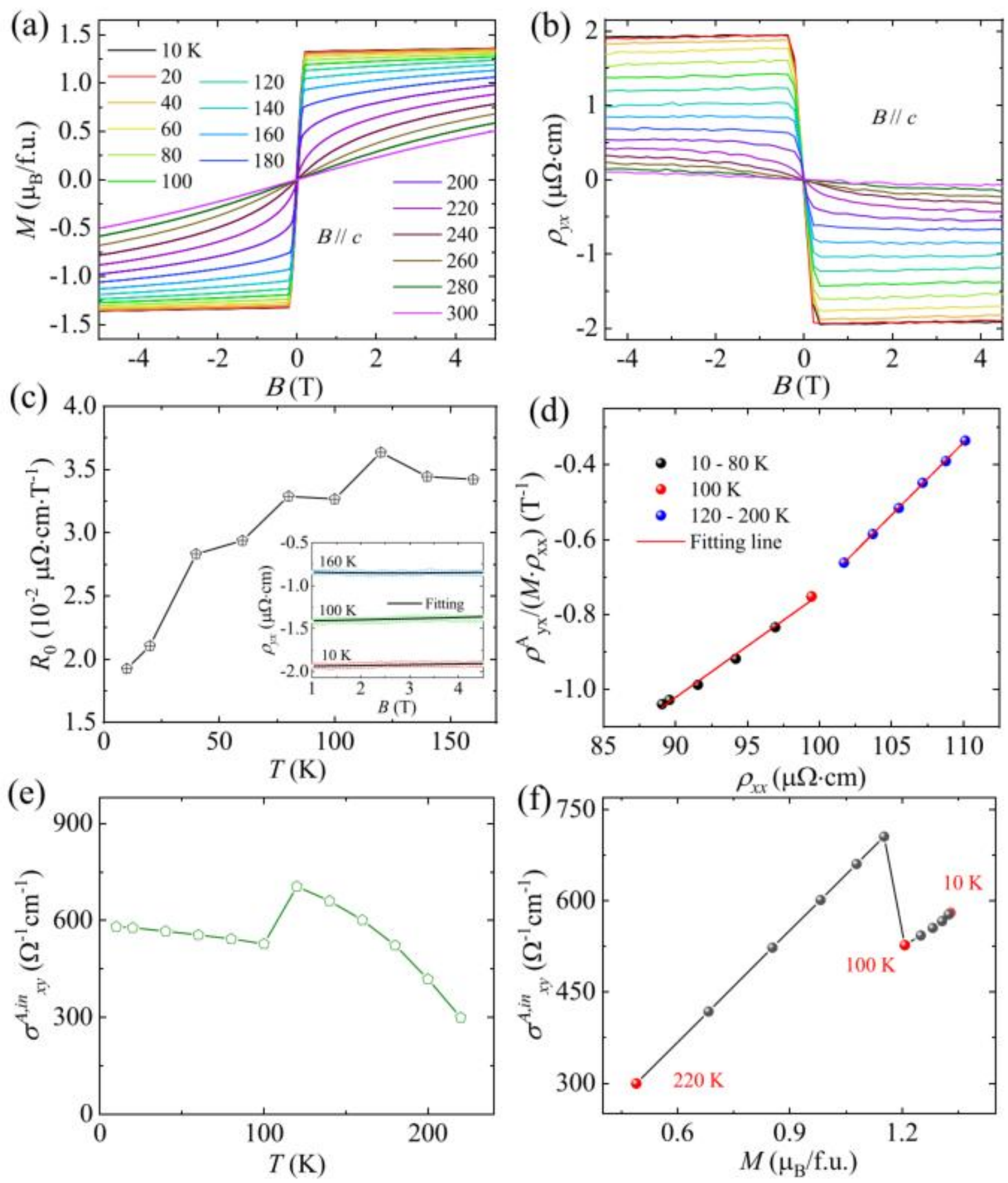

FIGURE 4 (a) Field-dependent magnetization with the field applied along *c*. (b) Field-dependent Hall resistivity $\rho_{yx}$. (c) Temperaturedependent ordinary Hall coefficient $R_0$. Inset: Plot of multiple linear regression fitting to the Hall resistivity at selected temperatures. (d) Plot of $\rho^A{}_{yx}/(M\rho_{xx})$ vs. $\rho_{xx}$. The data are taken at a magnetic field of 1 T. (e) $\sigma^{A,in}{}_{xy}$ as a function of temperature. (f) Plot of intrinsic AHC $\sigma^{A,in}{}_{xy}$ vs. *M*.

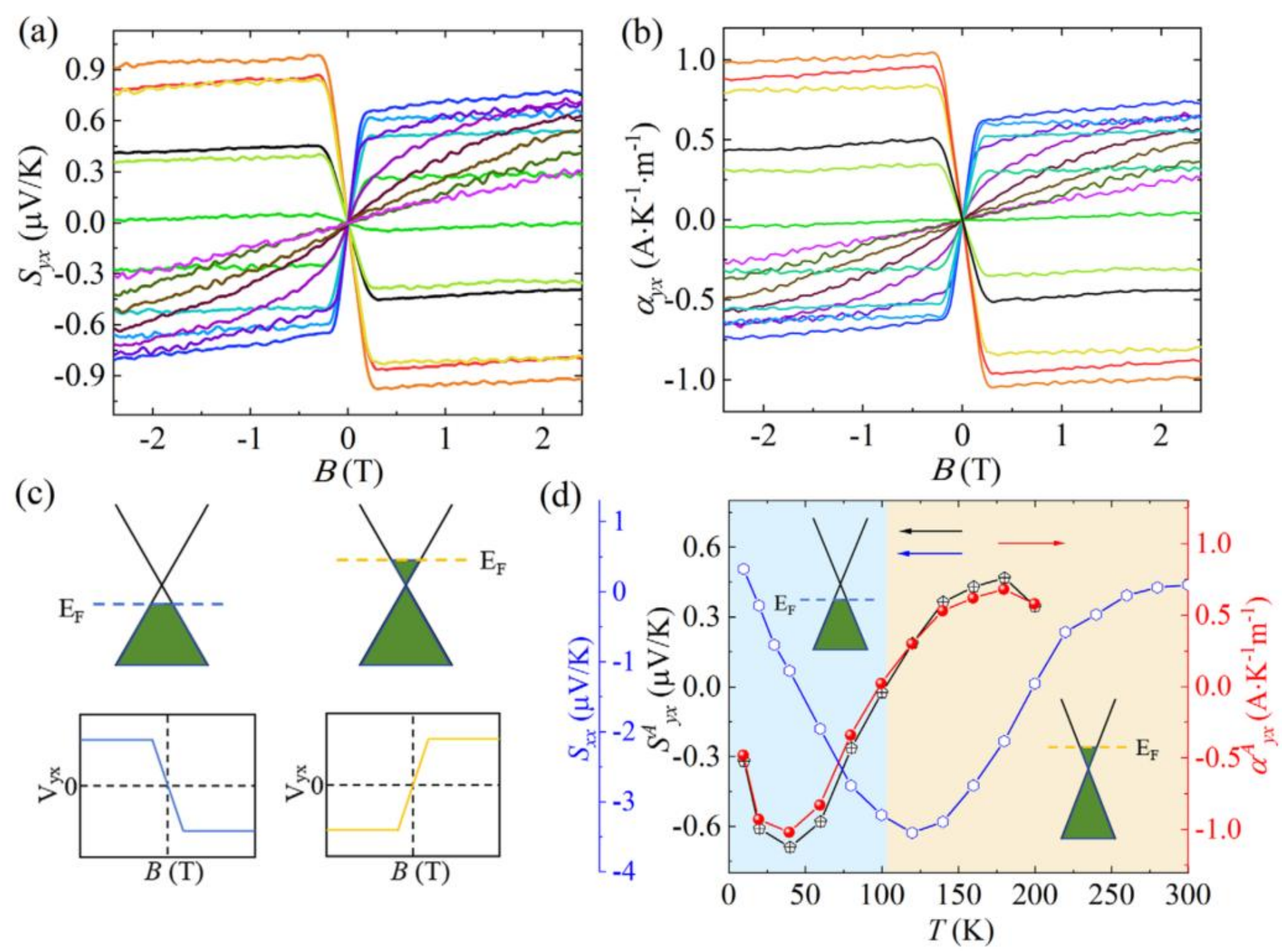

FIGURE 5 (a) Nernst coefficient as a function of magnetic field at various temperatures. The color scale representing temperature is consistent with that in Figure 4a. (b) Field-dependent Nernst conductivity $\alpha_{yx}$ at different temperatures. (c) Schematic diagram of the sign of the intrinsic transverse thermoelectric voltage in a Weyl magnet, which is determined by the relative position between the Fermi energy and the Weyl nodes. (d) Temperature evolution of the anomalous Nernst thermopower, anomalous Nernst conductivity, and Seebeck coefficient.

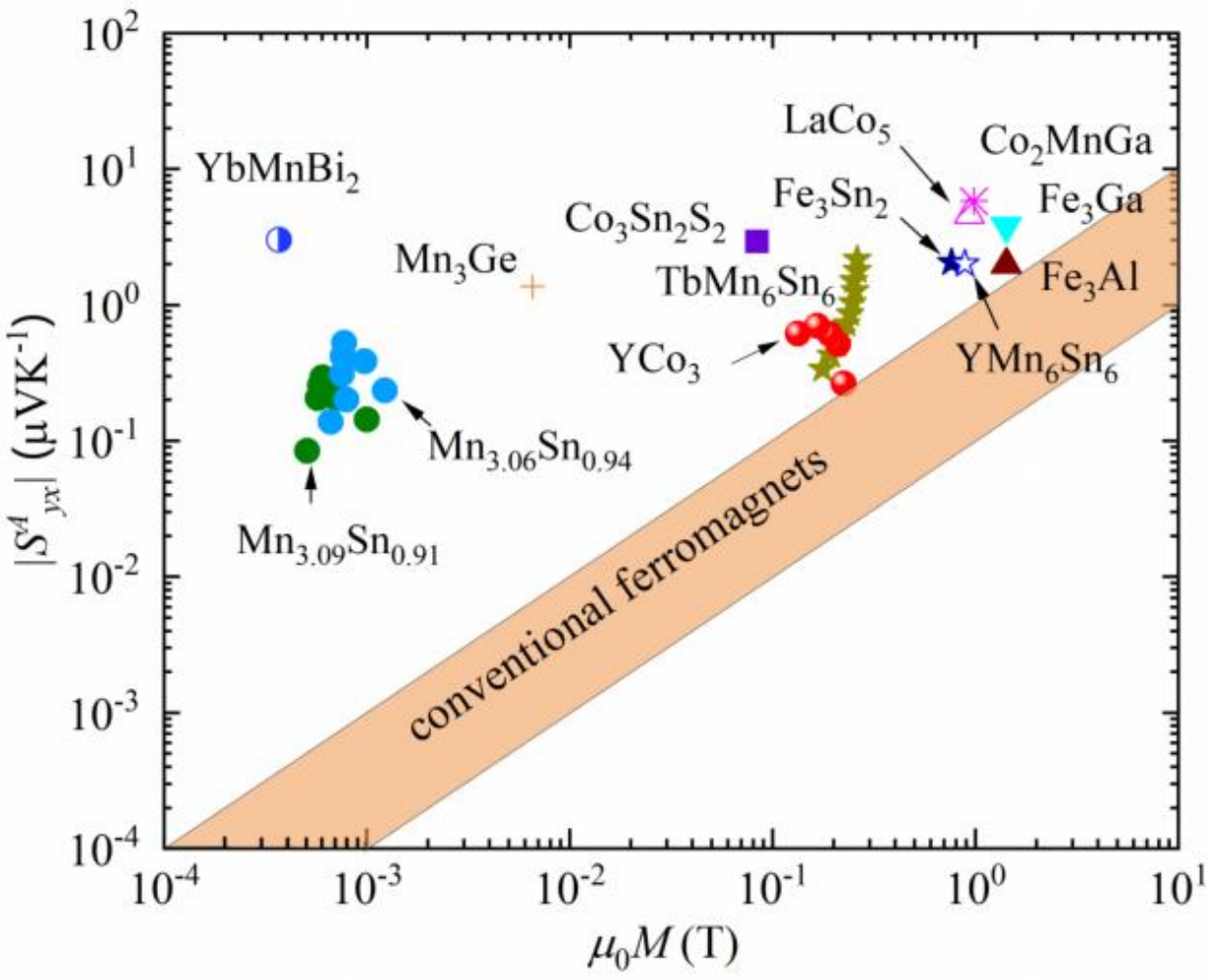

FIGURE 6 Scaling of $S^A_{yx}$ with $\mu_0 M$ of $YCo_3$ and other recently reported magnetic topological materials [13, 28, 30, 32, 33, 57–60].

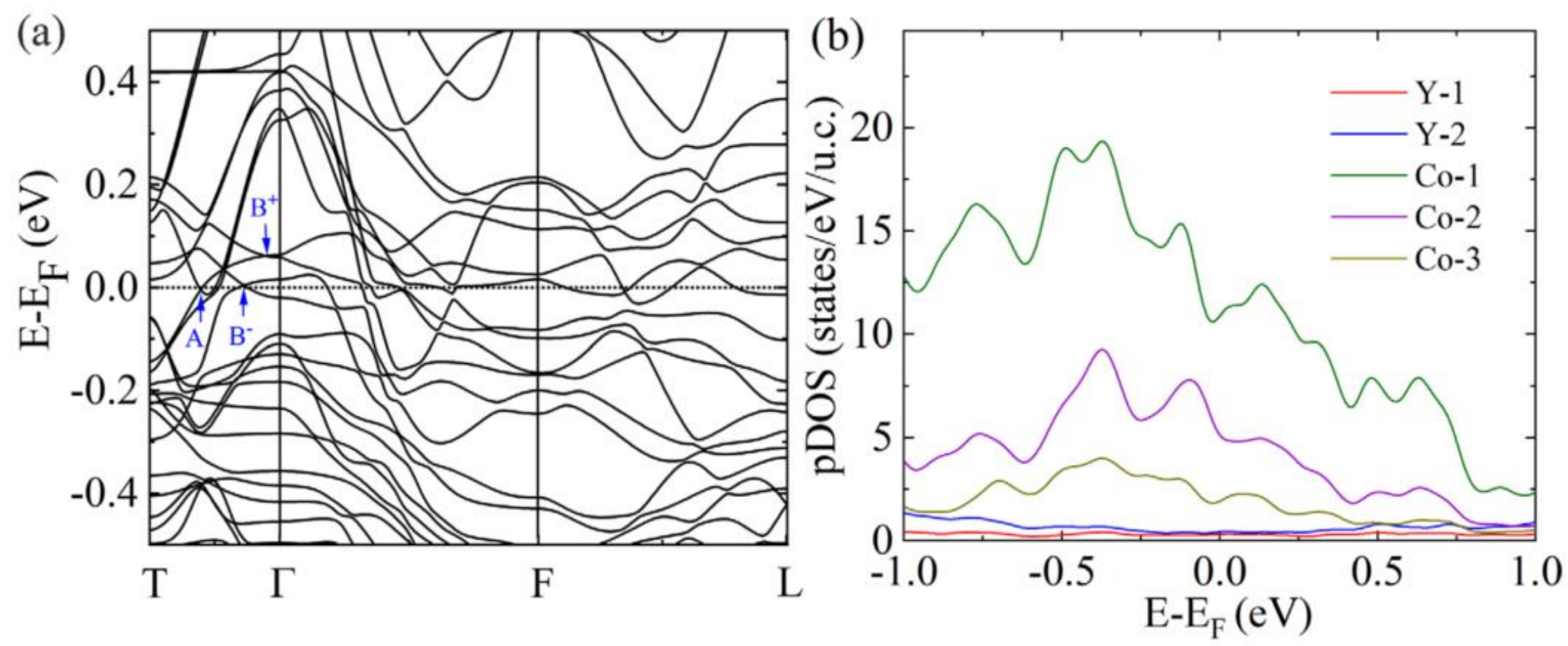

FIGURE 7 (a) The band structure of $YCo_3$ with SOC. (b) pDOS around the Fermi level.